\documentclass[twocolumn]{aastex62}

\newcommand{\GALEX}{{\it GALEX}}
\newcommand{\Gaia}{{\it Gaia}}

\newcommand{\HST}{{\it HST}}

\newcommand{\Spitzer}{{\it Spitzer}}
\newcommand{\Swift}{{\it Swift}}
\newcommand{\WISE}{{\it WISE}}

\newcommand{\Teff}{T_{\rm eff}}

\newcommand{\kms}{{\>\rm km\>s^{-1}}}

\newcommand{\bd}{BD+14$^\circ$3061}
\newcommand{\oCen}{$\omega$~Cen}
\newcommand{\uBVI}{{\it uBVI}}

\shorttitle{Spectra of Post-AGB Stars in M19}
\shortauthors{Bond et al.}

\begin{document}          

\title{Spectroscopic Confirmation of Two Luminous Post-AGB Stars in the Globular Cluster M19}

\correspondingauthor{Howard E. Bond}
\email{heb11@psu.edu}

\author[0000-0003-1377-7145]{Howard E. Bond}  
\affil{Department of Astronomy \& Astrophysics, Pennsylvania State
University, University Park, PA 16802, USA}
\affil{Space Telescope Science Institute, 3700 San Martin Drive,
Baltimore, MD 21218, USA}
\affil{Visiting Astronomer, Cerro Tololo Inter-American Observatory, National Optical Astronomy Observatory, operated by the Association of Universities for Research in Astronomy under a cooperative agreement with the National Science Foundation.}

\author[0000-0001-5754-4007]{Jacob E. Jencson}
\affil{Steward Observatory, 
University of Arizona, 
933 N Cherry Ave., 
Tucson, AZ 85721-0065, USA}

\author[0000-0002-1328-0211]{Robin Ciardullo}
\affil{Department of Astronomy \& Astrophysics,
Pennsylvania State University, University Park, PA 16802, USA}
\affil{Institute for Gravitation and the Cosmos, Pennsylvania
State University, University Park, PA 16802, USA}

\author[0000-0002-8994-6489]{Brian D. Davis}
\affil{Department of Astronomy \& Astrophysics, Pennsylvania
State University, University Park, PA 16802, USA}

\author[0000-0003-1817-3009]{Michael H. Siegel}
\affil{Department of Astronomy \& Astrophysics,
Pennsylvania
State University, University Park, PA 16802, USA}

\begin{abstract}

The visually brightest stars in globular clusters (GCs) are the ones evolving off the
asymptotic giant branch (AGB) and passing through spectral types F and A---the
``yellow'' post-AGB (yPAGB) stars. yPAGB stars are potentially excellent
``Population~II'' standard candles for measuring extragalactic distances. A
recent survey of the Galactic GC system, using \uBVI\/ photometry to identify
stars of low surface gravities with large Balmer discontinuities, discovered a
candidate luminous yPAGB star in the GC M19 (NGC~6273), designated ZNG\,4. The
same survey also identified a bright, hotter candidate blue PAGB star, ZNG\,2,
lying near the top of the white-dwarf cooling sequence. Both PAGB candidates
have proper motions and parallaxes in the recent \Gaia\/ Early Data Release~3 
consistent with cluster membership, but they still lacked spectroscopic
verification. Here we present moderate-resolution spectra of both stars,
confirming them as low-gravity objects that are extremely likely to be cluster
members. Through comparison with a library of synthetic spectra, we made
approximate estimates of the stars' atmospheric parameters. We find that the
yPAGB star ZNG\,4 has an effective temperature of $\Teff\simeq 6500$~K, a
surface gravity of $\log g\simeq1$, and a metallicity of $\rm[Fe/H]\simeq-1.5$,
similar to that of the host cluster. The blue PAGB star ZNG\,2 has
$\Teff\simeq18000$~K, $\log g\simeq3$, and an apparently low metallicity in the
range of $\rm[Fe/H]\simeq -2.0$ to $-2.5$. Both stars are bright ($V=12.5$ and
13.3, respectively). We urge high-dispersion spectroscopic follow-up to
determine detailed atmospheric parameters and chemical compositions, and to
monitor radial velocities.

\null\vskip0.2in

\end{abstract}


\section{Two Bright Post-AGB Candidates in~M19 \label{sec:intro} }

The most luminous objects in globular clusters (GCs) and other old stellar populations are stars that are in their final evolution from the top of the asymptotic giant branch (AGB) toward higher temperatures in the Hertzsprung-Russell diagram---the post-AGB (PAGB) stars. Evolving at nearly constant bolometric luminosity, PAGB stars attain their brightest visual magnitudes as they pass through spectral types F and A, where the bolometric correction is smallest. These luminous and non-variable ``yellow'' PAGB (yPAGB) stars have been proposed as potentially useful Population~II standard candles for measuring extragalactic distances, based on their expected narrow luminosity function in old populations, and ease of detection because of their large Balmer discontinuities \citep{Bond1997a, Bond1997b, Ciardullo2022}.

Due to the rapid evolutionary timescales of yPAGB stars, they are very rare objects. In a recent paper, \citet[][hereafter D22]{Davis2022} described their search of nearly the entire population of Galactic GCs for yPAGB stars, and other evolved stars lying above the horizontal branch (HB) in the clusters' color-magnitude diagrams (CMDs). The D22 survey was based on photometry in the \uBVI\/ system, which is optimized for detection of low-surface-gravity A-, F-, and G-type stars with large Balmer jumps (\citealt{Bond2005}; D22), combined with verification of cluster membership using astrometry from the recent \Gaia\/ Early Data Release~3 (EDR3; \citealt{Gaia2021}). 
As summarized by D22, only ten yPAGB stars are known in the Milky Way GC system, of which five are blueward of the Cepheid/RV~Tauri instability strip and are thus non-variable.

As PAGB objects evolve to still higher temperatures, they become ``blue'' PAGB (bPAGB) stars, potentially ionizing surrounding ejecta and becoming central stars of planetary nebulae. A recent review of bPAGB objects and other hot stars in GCs \citep{Moehler2019} lists about two dozen luminous bPAGB stars known in Galactic GCs. These stars then begin their descent of white-dwarf cooling tracks, at first fading rapidly to lower luminosities, and then cooling more slowly.

One of the results of the D22 survey was their discovery of a luminous yPAGB star, and a bright bPAGB star, in the
Galactic GC M19 (NGC~6273). These two stars were discussed in more detail by \citet[][hereafter B21]{Bond2021}. The yPAGB and bPAGB stars in M19 had been cataloged five decades ago as candidate ``UV-bright'' stars by \citet[][hereafter ZNG]{Zinn1972}, respectively designated M19 ZNG~4 and M19 ZNG~2. However, the \uBVI\/ study by B21 and D22 was the first to show that both objects have large Balmer jumps, consistent with the low surface gravities of luminous stars. This property distinguishes them from foreground stars, which are a substantial contaminant of the ZNG UV-bright catalog \citep[see][]{BondZNG2021}. Table~1 in B21 lists the coordinates, photometry, and other parameters of these two stars; that paper also presents a finding chart. 


Both of M19's PAGB stars have \Gaia\/ EDR3 proper motions and parallaxes that are consistent with cluster membership. But, without spectroscopic confirmation of their natures, they remained PAGB ``candidates,'' albeit very strong ones. In this paper, we discuss the evolutionary status of ZNG~4 and ZNG~2, and we then present spectroscopy of both stars, which strongly confirms them as luminous PAGB stars. This makes them worthy of further and more detailed study.

\section{Evolutionary Status \label{sec:evolstatus} }

\begin{figure*}
\begin{center}
\includegraphics[width=4.5in]{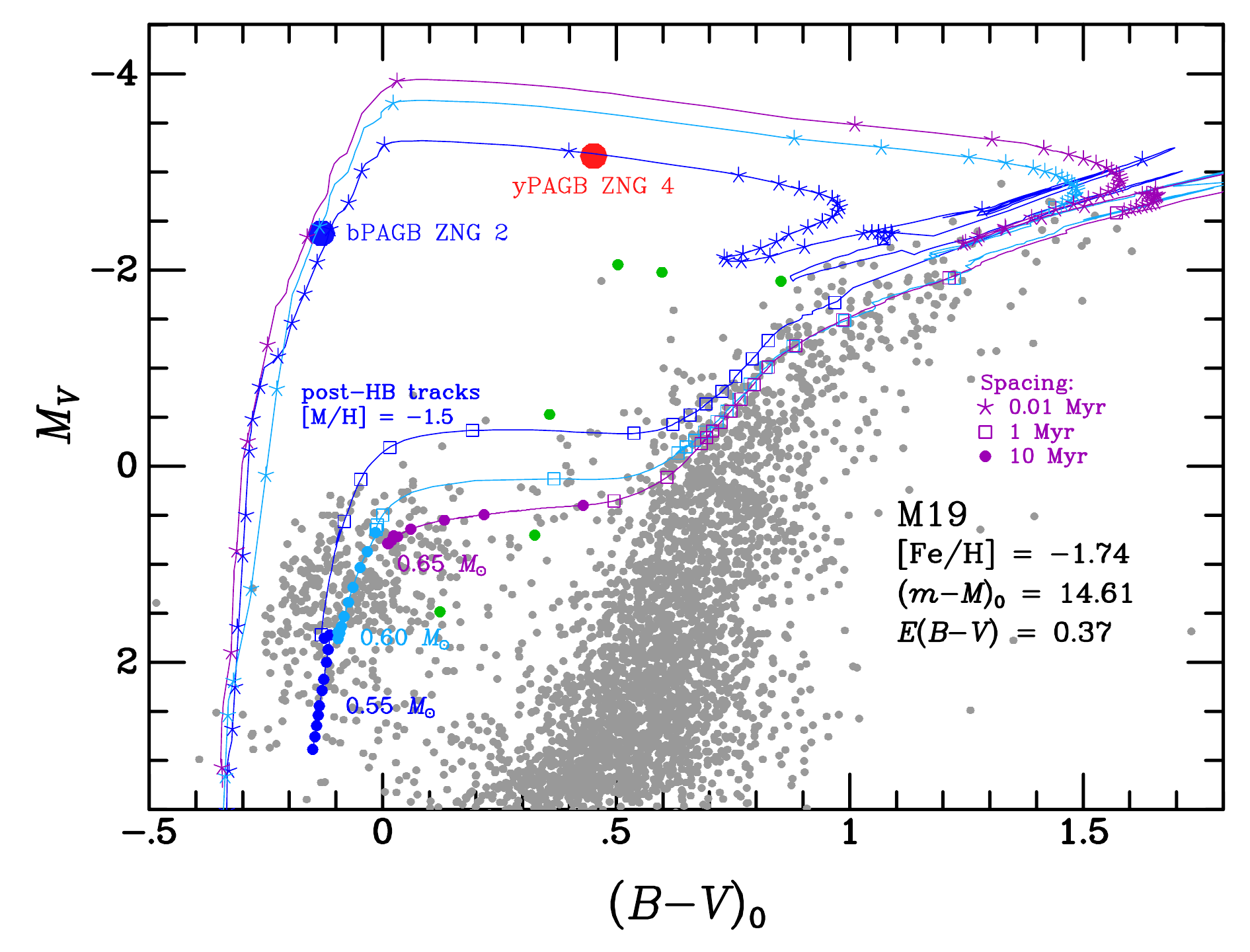} 
\figcaption{ 
Color-magnitude diagram for bright members of M19, based on data from \citet{Davis2022}, and converted to visual absolute magnitude and dereddened $B-V$ color, as described in the text and using parameters given in the figure legend. Filled gray circles are stars with high cluster membership probabilities based on \Gaia\/ proper motions and parallaxes, as described by \citet{Davis2022}. The two large filled circles mark the post-AGB stars ZNG\,2 and ZNG\,4. Small filled green circles plot the five known variable stars. Superposed are theoretical post-horizontal-branch evolutionary tracks from \citet{Moehler2019}, for a metallicity of $\rm[M/H]=-1.5$, and for the three zero-age-horizontal-branch masses indicated in the labels. The tracks have been converted to the observational quantities as described in the text. Time steps are marked on the tracks with the symbols given in the legend.
\label{fig:cmd_with_tracks}}
\end{center}
\end{figure*}

Figure~\ref{fig:cmd_with_tracks} presents the CMD [visual absolute magnitude, $M_V$, versus dereddened color, $(B-V)_0$] for a nearly complete catalog of bright members of M19. These data are from D22, and were obtained with CCD cameras on the 0.9- and 1.5-m telescopes at Cerro Tololo Inter-American Observatory. Field interloper stars---which are numerous at M19's low Galactic latitude of $b=+9\fdg4$---have been removed from the CMD based on EDR3 proper motions and parallaxes. We have retained only stars with membership probabilities greater than 0.8, as described in D22. The $BV$ data have been corrected for a mean cluster reddening of $E(B-V)=0.37$ \citep{Johnson2017}, and we have assumed $R_V=3.1$. M19 suffers from considerable differential extinction, which accounts for the broadening of the red-giant branch and HB in its CMD\null. For the two PAGB stars themselves, plotted in Figure~\ref{fig:cmd_with_tracks} as large filled circles, we obtained their reddenings from the detailed extinction maps of M19 given by \citet{Alonso2012} and \citet{Johnson2017}: $E(B-V)=0.344$ and 0.342 for ZNG~4 and ZNG~2, respectively. For the cluster distance, we adopt 8340~pc [$(m-M)_0=14.61$] from the recent compilation of Galactic GC distances assembled by \citet{Baumgardt2021}, which is based on \Gaia\/ parallaxes and other sources. This gives the stars visual absolute magnitudes of $M_V=-3.16$ and $-2.38$, respectively, with uncertainties of about $\pm$0.09~mag.

Superposed on the CMD in Figure~\ref{fig:cmd_with_tracks} are three theoretical post-HB evolutionary tracks computed by \citet{Moehler2019}.\footnote{We thank Marcelo Miller Bertolami for sending us tables of these tracks with a finer time resolution than given in the Moehler et al.\ paper.} As described in more detail in D22, we have edited and smoothed these tracks to remove several rapid excursions due to thermal pulses. We then converted the tracks from the theoretical quantities of luminosity and effective temperature to $M_V$ and $B-V$, using the online PARSEC YBC web tool\footnote{\url{http://stev.oapd.inaf.it/YBC/}} \citep{Chen2019}. These tracks are for a metal content of $\rm[M/H]=-1.5$. This is the nearest metallicity value in the Moehler et al.\ track library to that of M19. \citet{Harris2010} gives $\rm[Fe/H]=-1.74$ in his compilation of GC properties,\footnote{The Harris compilation of GC properties, 2010 December version, is available online at \url{http://physwww.mcmaster.ca/~harris/mwgc.dat}} 
while a spectroscopic study \citep{Johnson2017} of over 300 red giants and AGB stars in the cluster revealed a range of iron contents among the members, 
from about $\rm[Fe/H] = -2$ to $-1$, but concentrated around $-1.75$ and $-1.5$. 
Symbols along the tracks mark time steps of 10, 1, and 0.01~Myr, as indicated in the figure legend.

The HB in M19 is extremely blue, containing virtually no stars redder than $(B-V)_0\simeq0.15$. Filled green circles in Figure~\ref{fig:cmd_with_tracks} mark the D22 photometry of the five known variable stars in M19, which consist of only one RR~Lyrae variable, and four Type~II Cepheids.\footnote{As explained in D22, the data were obtained at a small number of epochs, and thus do not represent the mean locations of the variables in the CMD.} The three Moehler et al.\ tracks that are plotted are for masses on the zero-age HB (ZAHB) of 0.55, 0.60, and $0.65\,M_\odot$. These values encompass the range of ZAHB masses seen in this cluster. 

The yPAGB star ZNG\,4 lies close to the track for a ZAHB mass of $0.55\,M_\odot$. This is consistent with the finding discussed by D22 that all of the known yPAGB stars in Galactic GCs belong to clusters with blue horizontal branches. The explanation proposed by D22 is that this arises from the relatively slow evolutionary timescales on the PAGB for stars with low ZAHB masses. In this picture, stars with higher ZAHB masses do produce PAGB stars, but they evolve across the CMD extremely rapidly---as illustrated by the time marks on the tracks in Figure~\ref{fig:cmd_with_tracks}---and are thus so transient that none of them are present in the GC population.

The ZAHB mass of progenitor of the bPAGB star ZNG\,2 star is less certain, since the tracks at its location in the CMD are so close together. However, its ZAHB mass is statistically likely to be low, again because of the much more rapid evolutionary timescales for the more massive stars. In fact, it is remarkable that the two PAGB stars in M19 have nearly the same bolometric luminosity (see Table~1 in B21); their locations in the CMD are consistent with them both lying on nearly the same evolutionary track.


\section{Spectroscopic Observations and Reductions\label{sec:spectroscopy} }

Optical spectra of ZNG\,4 and ZNG\,2 were obtained on 2021 May~9 with the Boller \& Chivens spectrograph mounted on the 2.3-m Bok Telescope at the Steward Observatory station on Kitt Peak. A 1200~line\,mm$^{-1}$ grating was used in first order with a $1\farcs5$ slit, providing a wavelength coverage of 3520--4650~\AA\ at a 2.5-pixel resolution of 2.5~\AA\null. Exposure times were $3\times600$~s for ZNG\,4 and $3\times300$~s for ZNG\,2. HeArNe comparison-lamp spectra were obtained at the positions of both stars for wavelength calibration. M19 lies at a declination of $-26^\circ$, and the mean airmasses for the exposures were 2.0 and 2.3, respectively. The slit was oriented east-west, which unfortunately is roughly perpendicular to the parallactic angle.

We performed bias-subtraction, flat-fielding, and cosmic-ray removal on the CCD frames, using standard tasks in IRAF.\footnote{IRAF was distributed by the National Optical Astronomy Observatories, operated by AURA, Inc., under cooperative agreement with the National Science Foundation.} We then extracted the spectra from the CCD images, wavelength-calibrated them, and finally normalized them to a flat continuum, again employing standard IRAF tools. There is clear evidence of wavelength-dependent losses of flux due to atmospheric refraction, a consequence of the high airmasses and an east-west slit; therefore we did not attempt to flux-calibrate the spectra, nor measure radial velocities.



\section{Atmospheric Parameters \label{sec:parameters} }

Here we compare the spectra of the two PAGB stars with the extensive library of synthetic spectra computed by
\citet[][hereafter AP18]{AllendePrieto2018}, and available online from the Strasbourg Astronomical Data Center.\footnote{\url{http://cdsarc.u-strasbg.fr/viz-bin/qcat?J/A+A/618/A25}} These theoretical spectra are tabulated with a wavelength spacing of $\sim$0.13~\AA, and have a resolution of $\sim$0.4~\AA\null. For the purpose of comparison with our observations, we smoothed the synthetic spectra to a resolution of 2.5~\AA, using the {\tt boxcar} task in IRAF.

\subsection{Yellow PAGB Star ZNG 4}

\begin{figure*}[hb]
\begin{center}
\includegraphics[width=4.5in]{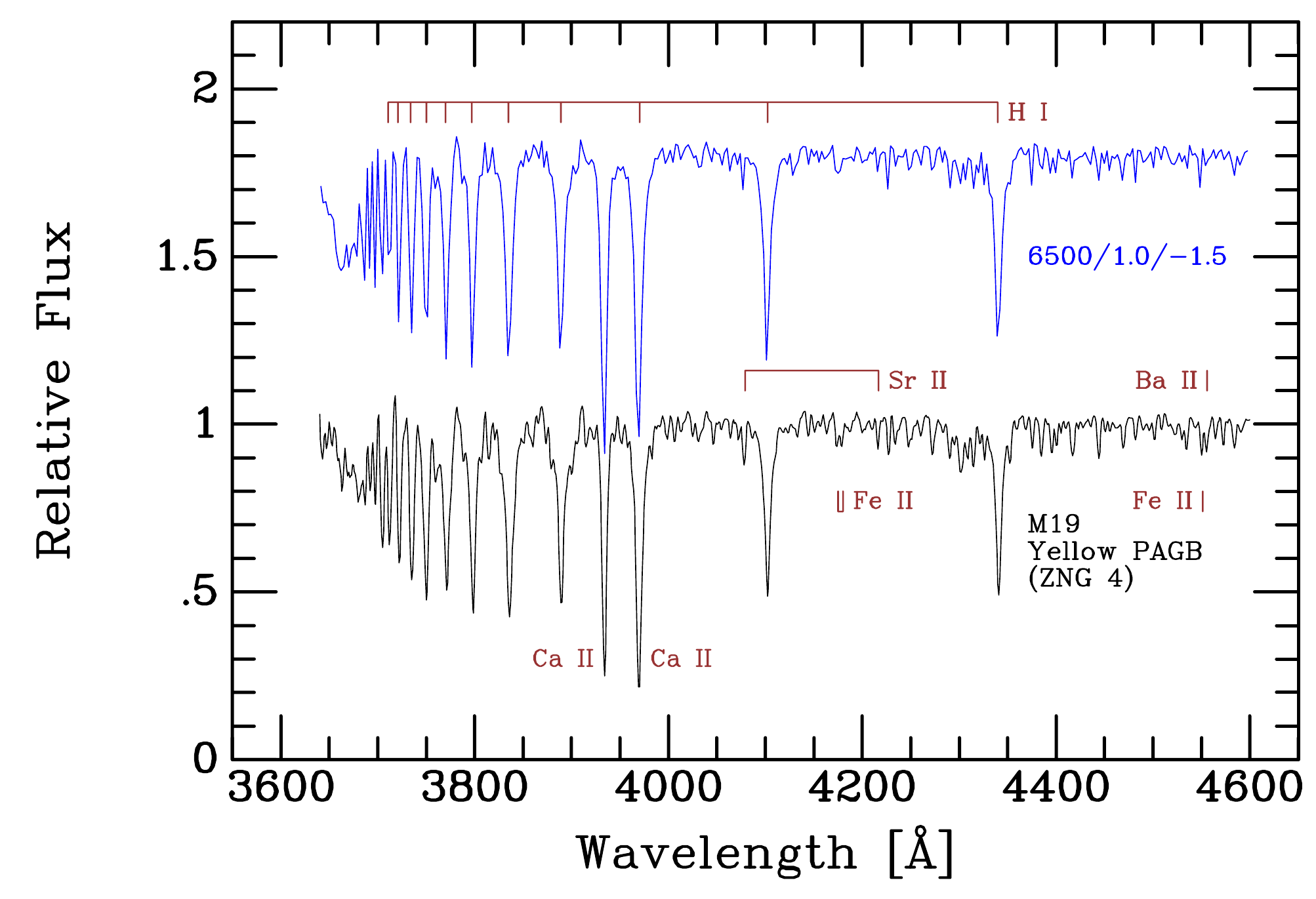} 
\figcaption{ 
Observed and model-atmosphere spectra for the yellow PAGB star M19 ZNG\,4, normalized to flat continua. Lines of several species are marked. {\it Black curve:} observed spectrum of ZNG\,4. Note the sharp and deep Balmer absorption lines and relatively strong lines of \ion{Fe}{2}, \ion{Sr}{2}, and \ion{Ba}{2}, indicative of a very low surface gravity. {\it Blue curve:} synthetic spectrum of a star with $\Teff=6500$~K, $\log g=1.0$, and $\rm[Fe/H]=-1.5$, shifted upward by 0.8 times the continuum level. It gives a good match to the observed spectrum.
\label{fig:yPAGB}}
\end{center}
\end{figure*}

The spectrum of ZNG\,4 is plotted as a black curve in Figure~\ref{fig:yPAGB}. The \uBVI\/ photometry (B21, D22) shows that ZNG\,4 has an extremely large Balmer discontinuity (B21 give a color of $u-B=1.47$), indicating a very low surface gravity. However, the normalization to a flat continuum, as shown in the figure, largely suppresses its depth when there is only a small amount of wavelength coverage below the jump. The spectrum of ZNG\,4 is clearly that of a very low-surface-gravity star, as indicated by its narrow and deep Balmer lines, and the relatively strong lines of ionized species such as \ion{Fe}{2}, \ion{Sr}{2}, and \ion{Ba}{2}, a few of which are marked in the figure.

To estimate the atmospheric parameters of the star, we compared its spectrum with a sequence of theoretical spectra from AP18 with a metallicity of $\rm[Fe/H]=-1.5$, surface gravity of $\log g=1.0$ (near the gravity anticipated from the star's location in the CMD), and effective temperatures with a spacing of 500~K\null. The closest match in this grid to the observed spectrum is for $\Teff=6500$~K, placing it blueward of the instability strip. The smoothed synthetic spectrum for this temperature is plotted as the blue curve in Figure~\ref{fig:yPAGB}. It provides a good match to the observed spectrum. (We also compared the spectrum to the grid at metallicity $\rm[Fe/H]=-2.0$, but the higher metal content gave a better fit to the metallic-line spectrum, as shown in the figure.)

 A number of yellow PAGB stars in the field \citep[e.g.,][and references therein]{Bond1991,Oomen2019}, as well as the yPAGB star in the GC M79 \citep{Sahin2009}, show photospheric depletions of refractory heavy elements. These depletions, which can be extreme \citep[e.g.,][]{Takeda2002,Rao2012}, are generally attributed to condensation onto grains which are then lost from the atmosphere. Yet ZNG\,4 does not appear to be significantly more metal-poor than its host cluster, based on our modest-resolution spectrum. A high-dispersion abundance analysis of ZNG\,4 would be of considerable interest, as would a radial-velocity study to search for binarity.

\subsection{Blue PAGB Star ZNG 2}

\begin{figure*}
\begin{center}
\includegraphics[width=4.5in]{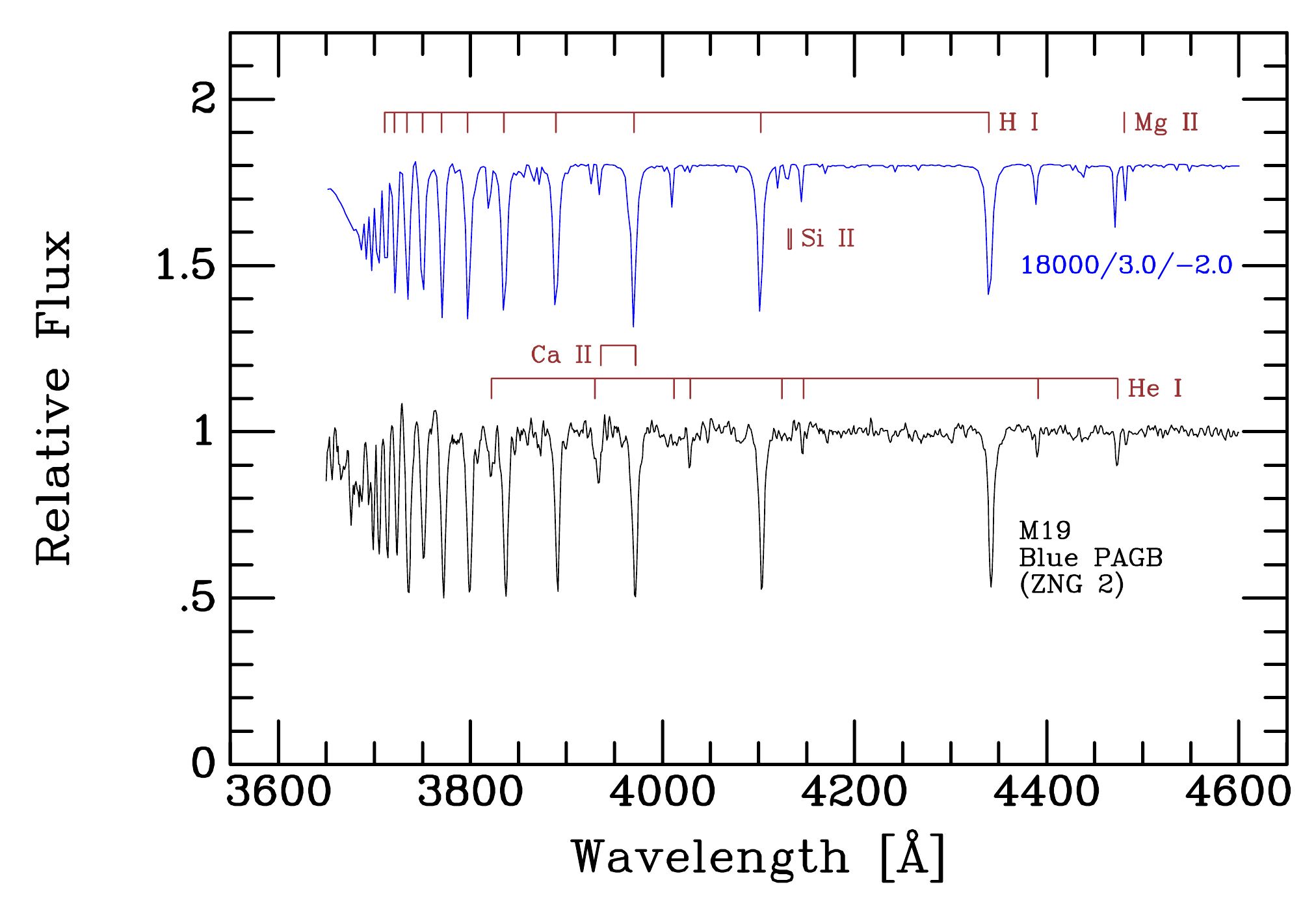} 
\figcaption{ 
Observed and model-atmosphere spectra for the blue PAGB star M19 ZNG\,2, normalized to flat continua. Lines of several species are marked. {\it Black curve:} observed spectrum of ZNG\,2. Note the sharp and deep Balmer absorption lines, indicative of a low surface gravity. {\it Blue curve:} synthetic spectrum of a star with $\Teff=18000$~K, $\log g=3.0$, and $\rm[Fe/H]=-2.0$, shifted upward by 0.8 times the continuum level. It gives a fairly good match to the observed spectrum, although the lines of \ion{Mg}{2} and \ion{Si}{2} in ZNG\,2 are slightly weaker than in the synthetic spectrum. The \ion{Ca}{2} lines in ZNG\,2 are likely mostly interstellar.
\label{fig:bPAGB}}
\end{center}
\end{figure*}

In Figure~\ref{fig:bPAGB} the spectrum of ZNG\,2 is shown as a black curve. As in the case of ZNG\,4, the normalization to a flat continuum has reduced the apparent depth of its Balmer discontinuity, which is actually large (B21 give a color of $u-B=0.61$). Numerous lines of \ion{He}{1} are present, indicative of a fairly high temperature.

As in the case of the yPAGB star, we compared this spectrum with spectra from the AP18 grid, assuming a surface gravity of $\log g=3.0$ (once again near the anticipated gravity). Spectra for $\rm[Fe/H]=-1.5$ gave metal lines consistently stronger than in ZNG\,2, so we adopted $\rm[Fe/H]=-2.0$. We then examined theoretical spectra to find the best fit to the overall strengths of the \ion{He}{1} lines, which proved to be for $\Teff=18,000$~K (the grid spacing is 2000~K at this temperature). This synthetic spectrum, smoothed to 2.5~\AA\ resolution, is plotted as the blue curve in Figure~\ref{fig:bPAGB}. It gives a fairly good match to the observed spectrum, although lines of \ion{Mg}{2} and \ion{Si}{2} are weaker in the star than in the model, and a few of the \ion{He}{1} lines are not well matched. (Conversely, the \ion{Ca}{2} K line is {\it stronger\/} in the star than in the model, but it is likely to have a strong interstellar component, given the relatively large reddening.) However, adopting models with $\rm[Fe/H]=-2.5$ would give lines of \ion{Mg}{2} and \ion{Si}{2} weaker than in the star, so the star's metallicity is likely in the range $-2.5<\rm[Fe/H]<-2.0$. There is thus some suggestion that ZNG\,2 is more metal-deficient than the red giants in M19, raising the possibility that depletion onto grains has occurred---but this suggestion needs to be confirmed with a high-dispersion abundance analysis.

\section{Conclusions}

We have obtained moderate-resolution spectra of the two candidate luminous PAGB stars in the Galactic GC M19 discovered by B21. We estimated the atmospheric parameters of the two stars through comparison with a grid of model-atmosphere synthetic spectra. 

The effective temperatures. low surface gravities, and low metallicities that we find are fully consistent with both stars being members of the cluster. Now that the cluster memberships of these very luminous and extremely rare stars appear to be firmly established, we urge high-dispersion spectroscopic studies of the their detailed chemical compositions, along with radial-velocity monitoring.

\acknowledgments


We thank Paul Smith for conducting the spectcroscopic observations presented in this work, and Nathan Smith for donating observing time.

This research has made use of the VizieR catalogue access tool, CDS,
Strasbourg, France (DOI: 10.26093/cds/vizier). The original description 
of the VizieR service was published in 2000, A\&AS 143, 23.

The Institute for Gravitation and the Cosmos is supported by the Eberly College of Science and the Office of the Senior Vice President for Research at The Pennsylvania State University.  

\facilities{Bok 2.3m; CTIO 0.9m, 1.5m}


\end{document}